\documentclass[preprint,aps,amsmath,superscriptaddress,nofootinbib]{revtex4}

\usepackage{dcolumn}
\usepackage{bm}
\usepackage{epsfig}
\usepackage{graphicx}
\usepackage{amsmath}
\usepackage{amssymb}
\usepackage{mathrsfs}
\usepackage{color}
\usepackage[usenames,dvipsnames]{xcolor}
\usepackage{hyperref}
\usepackage{braket}
\usepackage[caption=false]{subfig}

\usepackage{amsfonts}

\newcommand{\pv}{{\bf p}}

\newcommand{\lv}{{\bf l}}
\newcommand{\eps}{{\boldsymbol \epsilon}}

\newcommand{\ext}{{\rm ext}}

\def\slashchar#1{\setbox0=\hbox{$#1$}           
   \dimen0=\wd0                                 
   \setbox1=\hbox{/} \dimen1=\wd1               
   \ifdim\dimen0>\dimen1                        
      \rlap{\hbox to \dimen0{\hfil/\hfil}}      
      #1                                        
   \else                                        
      \rlap{\hbox to \dimen1{\hfil$#1$\hfil}}   
      /                                         
   \fi}

\begin{document}

\title{Strong decays of $T_{cc}^+$ at NLO in an effective field theory}


\author{Lin Dai}
\email{lin.dai@tum.de}
\affiliation{Physik Department, Technische Universit\"{a}t M\"{u}nchen, 85748 Garching, Germany\\}

\author{Sean Fleming}
\email{spf@email.arizona.edu}
\affiliation{Department of Physics and Astronomy, University of Arizona, Tucson, Arizona\ 85721, USA\\}

\author{Reed Hodges}
\email{reed.hodges@duke.edu}
\affiliation{Department of Physics, Duke University, Durham, North Carolina\ 27708, USA\\}

\author{Thomas Mehen}
\email{mehen@duke.edu}
\affiliation{Department of Physics, Duke University, Durham, North Carolina\ 27708, USA\\}

\preprint{TUM-EFT 173/22}

\begin{abstract} 
The $T_{cc}^+$ exotic meson, discovered  by the LHCb Collaboration in 2021, can be interpreted  as a molecular state of $D^{(*)0}$ and $D^{(*)+}$ mesons. We compute next-to-leading-order (NLO) contributions to the strong decay of $T_{cc}^+$ in an effective field theory for $D$ mesons and pions, considering contributions from one-pion exchange and final-state rescattering. Corrections to the total width, as well as the differential distribution in the invariant mass of the final-state $D$ meson pair are computed.  The results remain in good agreement with LHCb experimental results when the NLO contributions are added. The leading uncertainties in the calculation come from terms which depend on the scattering length and effective range in $D$ meson scattering. 
\end{abstract}

\maketitle

\section{Introduction}

The LHCb Collaboration has observed a narrow resonance, the exotic tetraquark $T_{cc}^+$, in the final-state $D^0 D^0 \pi^+$ \cite{Muheim,Polyakov,An,LHCb:2021vvq,LHCb:2021auc}. The resonance is close to both the $D^{*0}D^+$ and $D^{*+}D^0$ thresholds. When using a unitarized Breit-Wigner profile appropriate for a coupled-channel problem, LHCb finds the difference between the resonance mass and the $D^{*+} D^0$ threshold, $\delta m$, and the decay width, $\Gamma$, to be: \cite{LHCb:2021auc}
\begin{equation}\label{BW_width2}
\begin{aligned}
\delta m =& \; -360\pm40^{+4}_{-0}\, {\rm keV} \, ,  \\   
\Gamma  =& \;  48 \pm 2_{-14}^{+0}\, {\rm keV} \, .
\end{aligned}
\end{equation}
The $D^{*0}D^+$ threshold is 1.7 MeV above the resonance.  The closeness of the resonance to the two thresholds suggests the possibility that $T_{cc}^+$ has a molecular nature.  

After the announcement of the discovery of $T_{cc}^+$, many theory papers attempted to understand various aspects of the exotic meson \cite{Fleming:2021wmk,Meng:2021jnw,Agaev:2021vur,Wu:2021kbu,Ling:2021bir, Chen:2021vhg,Dong:2021bvy,Feijoo:2021ppq,Yan:2021wdl,Dai:2021wxi,Weng:2021hje,Huang:2021urd,Chen:2021kad,Xin:2021wcr,Albaladejo:2021vln,Du:2021zzh,Jin:2021cxj,Abreu:2021jwm,Dai:2021vgf,Deng:2021gnb,Azizi:2021aib}.   Several papers tried to predict its decay width and differential decay width, with considerable success \cite{Fleming:2021wmk,Meng:2021jnw,Ling:2021bir,Feijoo:2021ppq,Yan:2021wdl,Albaladejo:2021vln,Du:2021zzh}. In one of these papers \cite{Fleming:2021wmk}, we wrote down an effective field theory for $T_{cc}^+$ considering it a molecular state of two $D$ mesons treated nonrelativistically, and computed leading-order strong and electromagnetic decays.  Special attention was paid to the coupled-channel nature of the problem. We found a decay width of $52 \, {\rm keV}$ when the tetraquark is in an isospin 0 state, using a value of $\delta m=-273 \, {\rm keV}$, which arises from using a relativistic $P$-wave two-body Breit-Wigner function with a Blatt-Weisskopf form factor.  This was in good agreement with the LHCb experiment.  The predicted differential spectra as a function of the invariant mass of the final-state charm meson pair were also in good agreement with the binned experimental data. In this paper we investigate how these conclusions are affected by next-to-leading-order (NLO) strong decays.

The effective theory we will use is similar to the effective field theory for the $\chi_{c1}(3872)$ (XEFT) \cite{Fleming:2007rp,Fleming:2008yn,Fleming:2011xa,Mehen:2011ds,Margaryan:2013tta,Braaten:2010mg,Canham:2009zq,Jansen:2013cba,Jansen:2015lha,Mehen:2015efa,Alhakami:2015uea,Braaten:2015tga,Braaten:2020iye,Braaten:2020nmc,Braaten:2020iqw}. References \cite{Fleming:2007rp,Guo:2014hqa,Dai:2019hrf} have considered NLO XEFT diagrams for $\chi_{c1}(3872)$ decays.  One-pion exchange was found to have a negligible contribution to the decay width \cite{Fleming:2007rp,Dai:2019hrf}, while final-state rescattering led to uncertainty in the decay rate of ${}^{+50\%}_{-30\%}$ when the binding energy of the $\chi_{c1}(3872)$ is 0.2 MeV \cite{Dai:2019hrf}. The differential spectrum $d\Gamma[\chi_{c1}(3872)\rightarrow D^0 \bar{D}^0\pi^0]/dE_\pi$ was found to have a curve whose peak location and overall shape are insensitive to NLO corrections; only the normalization is affected \cite{Dai:2019hrf}.  The sharply peaked nature of the differential spectrum can inform about the molecular nature of the $\chi_{c1}(3872)$: since it is a function of the virtual $D^{*0}$ propagator $(p_D^2+\gamma^2)^{-1}$, where $\gamma$ is the binding momentum, as the binding energy goes to zero the distribution becomes sharply peaked as $p_D \to 0$. 

By analogy with this earlier work on $\chi_{c1}(3872)$, in this paper we compute NLO contributions to the decay of $T_{cc}^+$ to find the uncertainties due to one-loop one-pion exchange and final-state rescattering diagrams.  We calculate the uncertainty in the decay width, as well as in the shape, peak location, and normalization of differential spectra. The calculation is complicated by the presence of a coupled-channel, which is not present for $\chi_{c1}(3872)$. We find the decay width including NLO corrections to be $47_{-25\%}^{+53\%} \, {\rm keV}$, which is consistent with XEFT \cite{Dai:2019hrf}. We also discuss the physical significance of several of the parameters in the effective theory, and their effect on the decay width.

In Sec. \ref{secEffLag} we write down the effective Lagrangian to NLO. The required Feynman diagrams and their amplitudes, along with the explicit formulas for the partial widths are shown in Sec. \ref{secCoupChann}.  Plots of the differential distribution are shown in Sec. \ref{secPlots}, followed by concluding remarks in Sec. \ref{secConclusions}.

\section{Effective Lagrangian} \label{secEffLag}

The leading-order effective Lagrangian for strong decays of $T_{cc}^+$ is \cite{Fleming:2021wmk}
\begin{equation} \label{LagLO}
\begin{aligned}
 {\mathcal L}_{\rm LO} =& \; H^{* i\dagger}\bigg(i\partial^0+\frac{\nabla^2}{2m_{H^*}}-\delta^*\bigg)H^{* i} \\
 & + H^\dagger\bigg(i\partial^0+\frac{\nabla^2}{2m_H}-\delta\bigg)H  \\
& + \frac{g}{f_\pi} H^\dagger \partial^i \pi  H^{*i} + \text{H.c.} \\
&-C_0^{(0)} (H^{*T}\tau_2 H)^\dagger (H^{*T}\tau_2 H)\\
& - C_0^{(1)} (H^{*T}\tau_2 \tau_a H)^\dagger (H^{*T}\tau_2 \tau_a H) \, .
\end{aligned}
\end{equation}
Here $H$ and $H^*$ are isodoublets of the pseudoscalar and vector charm meson fields, respectively, and $\pi$ is the usual matrix of pion fields.  The diagonal matrices $\delta$ and $\delta^*$ contain the residual masses, which are the difference between the mass of the charm meson $D^{(*)i}$, where $i=0,+$, and that of the $D^0$. The coupling $g=0.54$ is the heavy hadron chiral perturbation theory (HH$\chi$PT) axial coupling \cite{Wise:1992hn,Burdman:1992gh,Yan:1992gz} and $f_\pi = 130$ MeV is the pion decay constant.  The terms on the last two lines are contact interactions mediating $D^*D$ scattering, where $C_0^{(n)}$ mediates $S$-wave scattering in the isospin-$n$ channel, and  $\tau_a$ are Pauli matrices acting in isospin space.

Several new classes of terms appear at NLO in the effective theory. There are new contact interactions involving two derivatives:
\begin{equation}\label{range}
\begin{aligned}
    {\mathcal L}_{C_2} &= \; \frac{C_2^{(0)}}{4} (H^{*T}\tau_2 H)^\dagger (H^{*T}\tau_2\overleftrightarrow{\nabla}^2 H) \\
    & + \frac{C_2^{(1)}}{4} (H^{*T}\tau_2 \tau_a H)^\dagger (H^{*T}\tau_2 \tau_a \overleftrightarrow{\nabla}^2H)  \\
    & + \, {\rm H.c.} \, .
\end{aligned}
\end{equation}
These interactions occur in XEFT and are proportional to the effective range \cite{Fleming:2007rp}. We can also write down $D\pi$ interaction terms by constructing isospin invariants out of the fields.
\begin{equation} \label{CpiLag}
\begin{aligned}
    {\mathcal L}_{C_\pi} =&\; C_\pi^{(1/2)} (\pi H)^\dagger (\pi H)  \\
    & + C_\pi^{(3/2)} \bigg(v_a H - \frac{1}{3}\tau_a\pi H \bigg)^\dagger  \\
    & \times \bigg(v_a H - \frac{1}{3}\tau_a\pi H \bigg) \, .
\end{aligned}
\end{equation}
Here $v = \begin{pmatrix} \pi^1 & \pi^2 & \pi^0 \end{pmatrix}^T/\sqrt{2}$ is a vector of pion fields, with $\pi^\pm \equiv (\pi^1 \mp i \pi^2)/\sqrt{2}$, such that $v_a \tau_a = \pi$.  $C_\pi^{(1/2)}$ and $C_\pi^{(3/2)}$ mediate scattering in the isospin-$1/2$ and isospin-$3/2$ channels, respectively. The interactions which are relevant to our calculation are: 
\begin{equation}
\begin{aligned}
    {\mathcal L}_{C_\pi} \rightarrow & \; C_\pi^{(1)}D^{0\dagger}\pi^{0\dagger} D^+ \pi^- - C_\pi^{(1)} D^{+\dagger}\pi^{0\dagger} D^0 \pi^+ + {\rm H.c.}   \\
    & + C_\pi^{(2)}D^{0\dagger}\pi^{0\dagger} D^0 \pi^0 + C_\pi^{(2)} D^{+\dagger}\pi^{0\dagger} D^+ \pi^0  \\
    & + C_\pi^{(3)}D^{0\dagger}\pi^{+\dagger} D^0 \pi^+ \; ,
\end{aligned}
\end{equation}
where the couplings $C_\pi^{(1)}$, $C_\pi^{(2)}$, and $C_\pi^{(3)}$ are particular linear combinations of $C_\pi^{(1/2)}$ and $C_\pi^{(3/2)}$ as governed by Eq.~(\ref{CpiLag}). These interactions can be matched onto the chiral Lagrangian \cite{Guo:2017jvc}. The values we use for these $C_\pi$ couplings are computed from lattice data; see Appendix \ref{CpiAppendix} for details.  

We can write down $D^*D\rightarrow DD\pi$ interactions by using the same strategy of constructing isospin invariants out of the fields.  That would lead to:
\begin{equation} \label{LagB11}
\begin{aligned}
    {\mathcal L}_{B_1} =&\; B_1^{(I=0)} \varepsilon_{\alpha \beta}(H^*_\alpha H_\beta)^\dagger(H \tau_2 \tau_iH \nabla v_i)  \\
    & + B_1^{(I=1)} (H^*\tau_2\tau_kH)^\dagger(\varepsilon_{ijk}H\tau_2\tau_iH \nabla v_j)  \\
    & + {\rm H.c.} \; .
\end{aligned}
\end{equation}
However, we need isospin-breaking terms in order to fully renormalize the theory at NLO, so ultimately we have four unique $B_1$ couplings, one for each possible channel.  Written in terms of the charm meson fields, the interactions become:
\begin{equation} \label{LagB12}
\begin{aligned}
    {\mathcal L}_{B_1} \rightarrow & \; B_1^{(1)}(D^+D^{*0})^\dagger(D^+D^0\nabla\pi^0)  \\
    & +B_1^{(2)}(D^0D^{*+})^\dagger(D^+D^0\nabla\pi^0) \\
    & +\frac{B_1^{(3)}}{2}(D^0D^{*+})^\dagger(D^0D^0\nabla\pi^+)  \\ 
    & + \frac{B_1^{(4)}}{2}(D^+D^{*0})^\dagger(D^0D^0\nabla\pi^+)  \\
    & + \, {\rm H.c.}\; .
\end{aligned}
\end{equation}
Relations between the $B_1^{(i)}$ implied by Eq.~(\ref{LagB11}) are given in the Appendix.
We can construct $DD$ contact terms out of the isospin invariants. There are only interactions in the isospin 1 channel,
\begin{equation}
\begin{aligned}
    {\mathcal L}_{C_{0D}} =&\; C_{0D}^{(1)}(H \tau_2 \tau_aH)^\dagger (H \tau_2 \tau_aH)  \\
    \rightarrow& \frac{C_{0D}^{(1)}}{2}(D^0D^0)^\dagger(D^0D^0)  \\
    & + C_{0D}^{(1)}(D^+ D^0)^\dagger (D^+ D^0) \, ,
\end{aligned}
\end{equation}
where in the second line we have restricted our focus to terms that are relevant to our calculation.
The authors in Ref.~\cite{Dai:2019hrf} chose to vary their $C_{0D}^{(1)}$ coupling, which described $D\bar{D}$ scattering as opposed to $DD$, over a range of $[-1,1]\,{\rm fm}^2$.  We test several different values for it within that range.  Lastly, we need a kinetic term for the pions;  in contrast to XEFT, we treat them relativistically,
\begin{equation}
\begin{aligned}
    {\mathcal L}_\pi ={\rm tr}(\partial^\mu\pi^\dagger \partial_\mu \pi - m_\pi ^2 \pi^\dagger\pi) \, .
\end{aligned}
\end{equation}
The full NLO Lagrangian is then ${\mathcal L}_{\rm NLO} = {\mathcal L}_{C_2} + {\mathcal L}_{C_\pi} + {\mathcal L}_{B_1} + {\mathcal L}_{C_{0D}} + {\mathcal L}_\pi$.

\begin{figure*}[t]
\centering
\includegraphics[trim=4.23cm 13.97cm 1.83cm 5.08cm,clip,scale=1]{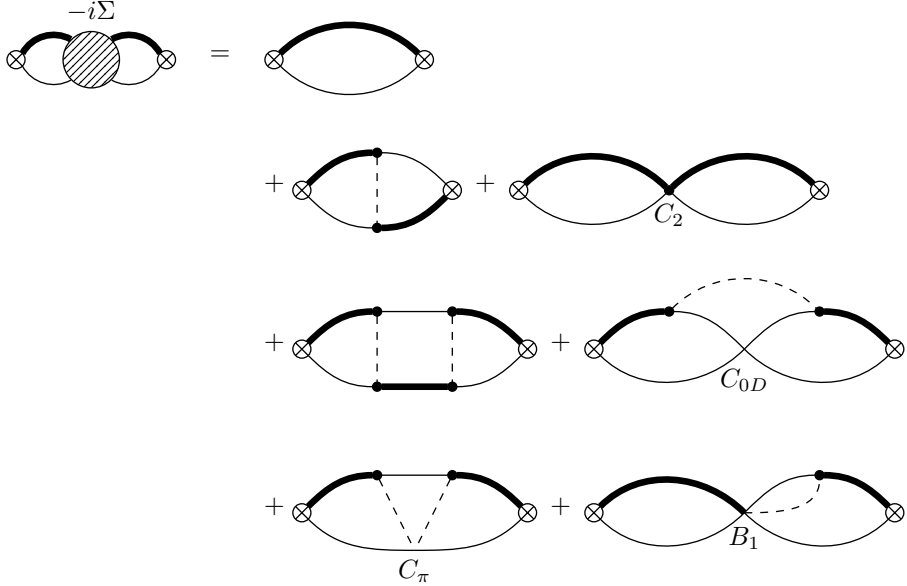}
\caption{Some of the $D^*D$ self-energy diagrams contributing to $-i\Sigma$.  Bold solid lines represent $D^*$ mesons, regular solid lines represent $D$ mesons, and dashed lines represent pions.  The first row is LO, the second row is NLO, and the third and fourth rows are NNLO.  There are also other NNLO diagrams not shown which are $C_0$-reducible combinations of the NLO diagrams.}
\label{bubbleDiagram}
\end{figure*}

\section{formulas for Decay Widths} \label{secCoupChann}

Writing down the decay width for the $T_{cc}^+$ at NLO requires care due to the coupled-channel nature of the problem.  We define a two-point correlation function matrix $\hat{G}$ as 
\begin{equation} \label{Geqn}
\begin{aligned}
\hat{G} =&\; \int d^4x\; e^{-iEt} \braket{0|T[X(x)X^T(0)]|0}  \\
& = i\Sigma(1+C\Sigma)^{-1} \, ,
\end{aligned}
\end{equation}
where the interpolating field is
\begin{equation}
\begin{aligned}
X=\begin{pmatrix} D^0D^{*+} \\ D^+D^{*0} \end{pmatrix} \, .
\end{aligned}
\end{equation}
The right-hand side of Eq.~(\ref{Geqn}) arises from expressing $\hat{G}$ to all orders as an infinite sum of the $C_0$-irreducible two-point function $\Sigma$, in a manner similar to that in Appendix A of Ref.~\cite{Kaplan:1998sz}, but here $C_0$ and $\Sigma$ are matrices due to the presence of a coupled-channel.  $-i\Sigma$ is given by the sum of $D^*D$ self-energy diagrams in Fig. \ref{bubbleDiagram}.  Its diagonal elements correspond to those two-point diagrams which do not swap channels, and the off-diagonal elements to those which do swap channels.  We can then project out the isospin 0 and isospin 1 channels, and tune the parameters of the two-point correlators so that there is a pole corresponding to the location of the $T_{cc}^+$ bound state. Near the vicinity of the pole, the Green's function can be written as 
\begin{equation} \label{G01}
\begin{aligned}
G_{0/1} =&\; \left(\begin{array}{c} 1 \\ \mp1 \end{array}\right)^T\hat{G}\left(\begin{array}{c} 1 \\ \mp1 \end{array}\right)  \\
 \approx & \frac{1}{2}\frac{iZ_{0/1}}{E+E_T + \frac{i\Gamma_{0/1}}{2}} \, ,
\end{aligned}
\end{equation}
where
$\Gamma_{0/1}$ is the decay width and the residue $Z_{0/1}$ is the wave function renormalization.   We find for the decay width in the isospin 0 channel
\begin{equation} \label{width}
\begin{aligned}
\Gamma_0^{NLO} \approx & \; -\Gamma^{LO}\frac{{\rm Re}\, \Sigma_0^{\prime NLO}(-E_T)}{{\rm Re}\, {\rm tr}\,\Sigma^{\prime LO}(-E_T)}  \\
& + \frac{2\,{\rm Im}\, \Sigma_0^{NLO}(-E_T)}{{\rm Re}\, {\rm tr}\,\Sigma^{\prime LO}(-E_T)} \, ,
\end{aligned}
\end{equation}
where $\Sigma_0 \equiv \Sigma_{11}+\Sigma_{22}-\Sigma_{12}-\Sigma_{21}$ is a particular combination of the elements of the $\Sigma$ matrix appropriate for isospin 0. The first term of Eq.~(\ref{width}) is a correction to the LO decay width from NLO $D^*D$ self-energy corrections, i.e., diagrams on the second row of Fig. \ref{bubbleDiagram}.  The second term of Eq.~(\ref{width}) consists of NLO decay diagrams, from various cuts of diagrams on the third and fourth rows of Fig. \ref{bubbleDiagram}. Note that ${\rm Im}\,\Sigma^{NLO}$ is from $\Sigma$ diagrams of one higher order than in ${\rm Re}\,\Sigma^{NLO}$ because the LO self-energy graph has no imaginary part below threshold.  The derivatives of $\Sigma$ are with respect to $E$ and evaluated at $E=-E_T$.  For a more detailed derivation of Eq.~(\ref{width}) refer to Appendix \ref{CoupledChannelAppendix}.


Three diagrams in Fig.~\ref{bubbleDiagram} contribute to ${\rm Re} \, \Sigma$ to NLO.  They are the LO self-energy diagram ($-i\Sigma_1$), the one-pion exchange diagram ($-i\Sigma_2$), and the $C_2$ contact diagram ($-i\Sigma_3$). They are evaluated in the power divergence subtraction (PDS) scheme \cite{Kaplan:1998tg}. 
This scheme corresponds to using $\overline{{\rm MS}}$ to handle logarithmic divergences as well as subtracting poles in $d=3$ to keep track of linear divergences.
A $1/\epsilon$ pole appears in $\Sigma_2$, but the dependence on the renormalization scale drops out when the derivative with respect to $E$ is taken.  We neglect terms in the propagators that go as $\pv^4/m_H^2$ or $(\delta m)\pv^2/m_H$, where $\delta m$ is of the order of the pion mass, compared to $\pv^2$.  In $\Sigma_2$ and $\Sigma_3$ we use a Fourier transform to evaluate the integrals over three-momentum, using a procedure outlined in Ref.~\cite{Braaten:1995cm}.   We define a reduced mass $\mu(m_1,m_2)\equiv m_1 m_2/(m_1+m_2)$ and the binding momenta are defined to be $\gamma^2(m_1,m_2)=2\mu(m_1,m_2)(m_1+m_2-m_T)$. The expressions for the self energy diagrams are:
\begin{widetext}
\begin{equation}
\begin{aligned}
-i\Sigma_1(m,m^*) =&\; -\frac{i\mu(m,m^*)}{2\pi}[\Lambda_{\rm PDS}-\gamma(m,m^*)] \, , \\
\end{aligned}
\end{equation}
\begin{equation}
\begin{aligned}
-i\Sigma_2(m_1,m_1^*,m_2,m_2^*,m_\pi,g_1,g_2) =&\; -\frac{4ig_1 g_2}{3}\mu(m_1,m_1^*)\mu(m_2,m_2^*)  \\
& \times \bigg\{\frac{1}{16\pi^2}[\Lambda_{\rm PDS}-\gamma(m_1,m_1^*)][\Lambda_{\rm PDS}-\gamma(m_2,m_2^*)]  \\
& + \frac{(m_2^*-m_1)^2-m_\pi^2}{(8\pi)^2} \bigg[\frac{1}{\epsilon}+2  \\
& -4 \log \bigg( \gamma(m_1,m_1^*)+\gamma(m_2,m_2^*)  \\
& -i(m_2^*-m_1)^2+im_\pi^2 \bigg) -4\log \mu \bigg] \bigg\} \, , \\
\end{aligned}
\end{equation}
\begin{equation}
\begin{aligned}
-i\Sigma_3(m_1,m_1^*,m_2,m_2^*,C_2) =&\; -\frac{i}{4\pi^2}C_2[\gamma^2(m_1,m_1^*)+\gamma^2(m_2,m_2^*)]\mu(m_1,m_1^*) \\ &  \times \mu(m_2,m_2^*) 
 [\Lambda_{\rm PDS}-\gamma(m_1,m_1^*)][\Lambda_{\rm PDS}-\gamma(m_2,m_2^*)] \, .
\end{aligned}
\end{equation}
\end{widetext}
\begin{figure*}[t]
\centering
\begin{minipage}{0.33\textwidth}
\centering
\subfloat[]{\includegraphics[trim=5.08cm 19.05cm 11.43cm 3.81cm,clip,scale=0.8]{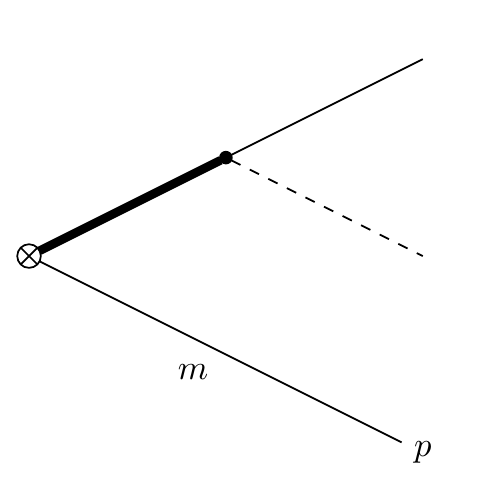}\label{figX1a}}
\end{minipage}%
\begin{minipage}{0.33\textwidth}
\centering
\subfloat[]{\includegraphics[trim=5.08cm 19.05cm 11.43cm 3.81cm,clip,scale=0.8]{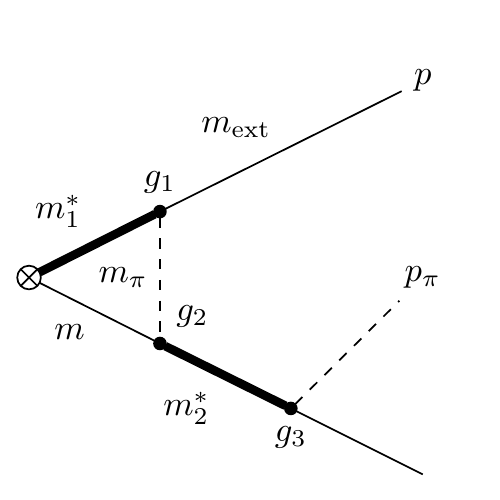}\label{figX1b}}
\end{minipage}%
\begin{minipage}{0.33\textwidth}
\centering
\subfloat[]{\includegraphics[trim=5.08cm 19.05cm 9.74cm 3.81cm,clip,scale=0.8]{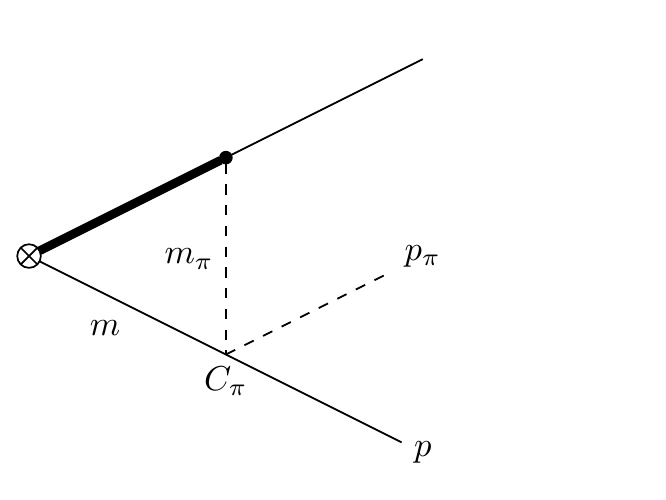}\label{figX1d}}
\end{minipage}
\bigskip
\begin{minipage}{0.33\textwidth}
\centering
\subfloat[]{\includegraphics[trim=5.08cm 19.05cm 9.74cm 3.81cm,clip,scale=0.8]{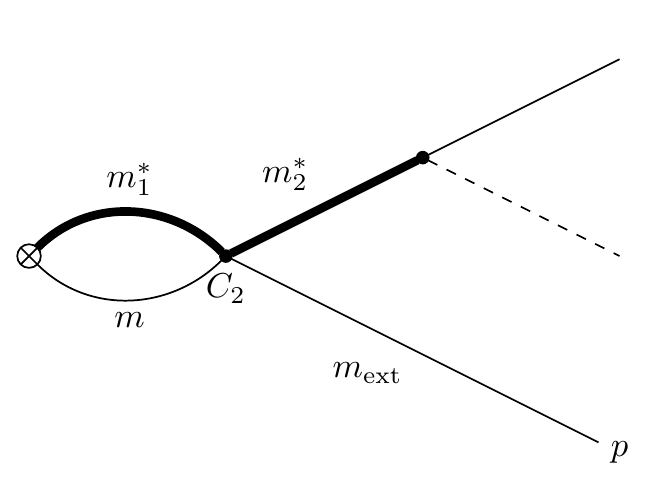}\label{figX1e}}
\end{minipage}%
\begin{minipage}{0.33\textwidth}
\centering
\subfloat[]{\includegraphics[trim=5.08cm 21.17cm 11.43cm 3.81cm,clip,scale=0.8]{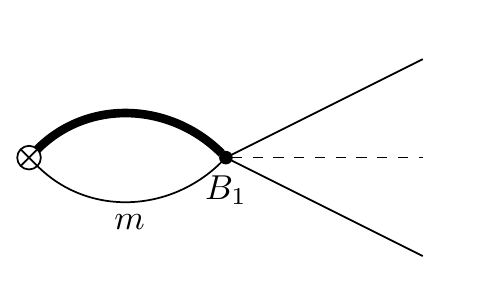}\label{figX1f}}
\end{minipage}%
\begin{minipage}{0.33\textwidth}
\centering
\subfloat[]{\includegraphics[trim=5.08cm 21.17cm 11.43cm 3.81cm,clip,scale=0.8]{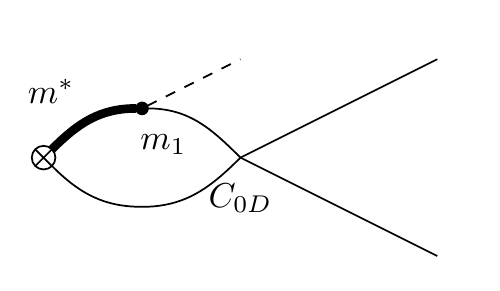}\label{figX1g}}
\end{minipage}
\caption{Feynman diagrams at LO and NLO contributing to the decay of $T_{cc}^+$.  We label the vertices and lines whose naming might be ambiguous.  These diagrams arise from cuts of the diagrams on the third and fourth lines of Fig. \ref{bubbleDiagram}.}
\label{DecayDiagrams}
\end{figure*}
To be consistent with the implementation of the PDS scheme in the decay diagrams (see Appendix \ref{BasisInt}), for the double integral in $\Sigma_2$ we have used rotational symmetry to replace the tensor structure in the numerator with $\delta^{ij}/3$ and not $\delta^{ij}/(d-1)$.  This choice does not affect the derivative of $\Sigma_2$ as it only changes the constant  terms which drop out upon differentiation with respect to $E$.

The decay diagrams that contribute to $2\,{\rm Im}\, \Sigma_0^{NLO}(-E_T)$ are shown in Fig.~\ref{DecayDiagrams}.   By the optical theorem the square of these diagrams is given by the sum over the cuts of the NNLO diagrams in Fig.~\ref{bubbleDiagram}. If there is only one pion/charm meson vertex in a diagram, its coupling is labeled $g_\pi$.  If there are more than one such vertex, the couplings are numbered $g_i$.  Depending on the type of pion and charm meson, these couplings will be either $g/f_\pi$ or $\pm g/(\sqrt{2}f_\pi$).The expressions are written in terms of the basis integrals given in Appendix \ref{BasisInt}. These basis integrals depend on parameters $b$, $c_1$, and $c_2$, the definitions for $c_1$ and $c_2$ are provided where appropriate, $b=1$ unless otherwise specified, and the momentum arguments for the integrals are $\pv$ unless otherwise specified.   

\begin{widetext}

\begin{equation}
\begin{aligned}
i\mathcal{A}_{{\rm(\ref{figX1a})}}(\pv,m,m^*,g_\pi) =&\; \frac{2ig_\pi \eps_T \cdot \pv_\pi \mu(m,m^*)}{\pv^2+\gamma^2(m,m^*)} \, . \\
\end{aligned}
\end{equation}
\begin{equation}
\begin{aligned}
i\mathcal{A}_{{\rm(\ref{figX1b})}}(\pv,m,m_{\rm ext},m_\pi,m_1^*,m_2^*,g_1,g_2,g_3) =&\;  \frac{4i \mu(m,m_1^*)\mu(m_{\rm ext},m_2^*)g_1g_2g_3}{\pv^2+\gamma^2(m_{\rm ext},m_2^*)}  \\
& \times \bigg[ \eps_T\cdot\pv \,\pv_\pi\cdot\pv\big(I_0^{(2)}-2I^{(1)}+I\big)  \\
&+\eps_T\cdot\pv_\pi\pv^2 I_1^{(2)}\bigg] \, ,  \\
c_1 =&\; \gamma^2(m,m_1^*)\, ,  \\
c_2 =&\; \pv^2-(m_T-m-m_\ext)^2+m_\pi^2 \, .  \\
\end{aligned}
\end{equation}
\begin{equation}
\begin{aligned}
i\mathcal{A}_{{\rm(\ref{figX1d})}}(m,m_{\rm ext},m_\pi,m^*,g_\pi,C_\pi) =&\; 2i\mu(m,m^*)g_\pi C_\pi \eps_T\cdot\pv[I^{(1)}-I] \, , \\
c_1 =&\; \gamma^2(m,m^*)\, ,  \\
c_2 =&\; \pv^2-(m_T - m-m_\ext)^2+m_\pi^2 \, .  \\
\end{aligned}
\end{equation}
\begin{equation}
\begin{aligned}
i\mathcal{A}_{{\rm(\ref{figX1e})}}(m,m_\ext,m_1^*,m_2^*,g_\pi,C_2) =&\; \frac{1}{\pi}iC_2g_\pi\eps_T\cdot\pv_\pi\mu(m,m_1^*)\mu(m_{\rm ext},m_2^*)  \\
&\times \frac{\pv^2-\gamma^2(m,m_1^*)}{\pv^2+\gamma^2(m_\ext,m_2^*)} [\gamma(m,m_1^*)-\Lambda_{\rm PDS}] \, . \\
\end{aligned}
\end{equation}
\begin{equation}
\begin{aligned}
i\mathcal{A}_{{\rm(\ref{figX1f})}}(m,m^*,B_1) =&\; -\frac{iB_1}{2\pi} \eps_T\cdot\pv_\pi \mu(m,m^*)[\gamma(m,m^*)-\Lambda_{\rm PDS}] \, . \\
\end{aligned}
\end{equation}
\begin{equation}
\begin{aligned}
i\mathcal{A}_{{\rm(\ref{figX1g})}}(m_1,m_2,m^*,p_\pi^0,g_\pi,C_{0D}) =&\; 4i\mu(m_1,m_2)\mu(m_2,m^*) g_\pi C_{0D}\eps_T\cdot\pv_\pi I(\pv_\pi) \, , \\
c_1 =&\; \gamma^2(m_2,m^*) \, ,  \\
c_2 =&\; -2\mu(m_1,m_2)\bigg(m_T-m_1-m_2 - p_\pi^0 - \frac{\pv_\pi^2}{2m_1}\bigg) \, ,  \\
b =&\; \frac{\mu(m_1,m_2)}{m_1} \, . 
\end{aligned}
\end{equation}

\end{widetext}

Following Eq.~(\ref{width}) and using the amplitudes defined above, the decay widths for the two strong decays of $T_{cc}^+$ are

\begin{widetext}
\begin{equation}
\begin{aligned}
    \frac{d\Gamma_0^{NLO}(T_{cc}^+\rightarrow D^+D^0\pi^0)}{d\pv_0^2d\pv_+^2} =&\; \frac{2}{{\rm Re}\,{\rm tr} \, \Sigma^{\prime LO}(-E_T)}{\rm Re}\, \bigg[ {\mathcal A}_{(\ref{figX1a})}(\pv_+,m_+,m_0^*,-g/\sqrt{2}f_\pi)  \\
    & \times \bigg( \mathcal{A}_{(\ref{figX1b})}(\pv_0,m_+,m_0,m_{\pi^0},m_0^*,m_+^*,-g/\sqrt{2}f_\pi,g/\sqrt{2}f_\pi,g/\sqrt{2}f_\pi)  \\
    & + {\mathcal A}_{(\ref{figX1b})}(\pv_+,m_+,m_+,m_{\pi^-},m_0^*,m_0^*,g/f_\pi,g/f_\pi,-g/\sqrt{2}f_\pi)  \\
    & - {\mathcal A}_{(\ref{figX1b})}(\pv_0,m_0,m_0,m_{\pi^+},m_+^*,m_+^*,g/f_\pi,g/f_\pi,g/\sqrt{2}f_\pi)  \\
    & - {\mathcal A}_{(\ref{figX1b})}(\pv_+,m_0,m_+,m_{\pi^0},m_+^*,m_0^*,g/\sqrt{2}f_\pi,-g/\sqrt{2}f_\pi,-g/\sqrt{2}f_\pi)  \\
    & + {\mathcal A}_{(\ref{figX1d})}(\pv_0,m_+,m_0,m_{\pi^0},m_0^*,-g/\sqrt{2}f_\pi,C_\pi^{(2)})  \\
    & - {\mathcal A}_{(\ref{figX1d})}(\pv_0,m_0,m_0,m_{\pi^+},m_+^*,g/f_\pi,C_\pi^{(1)})  \\
    & +{\mathcal A}_{(\ref{figX1g})}(m_0,m_+,m_0^*,-g/\sqrt{2}f_\pi,C_{0D}^{(1)})  \\
    & -{\mathcal A}_{(\ref{figX1g})}(m_+,m_0,m_+^*,g/\sqrt{2}f_\pi,C_{0D}^{(1)}) \bigg)^* + (D^0 \leftrightarrow D^+, \pi^+ \leftrightarrow \pi^-) \bigg]  \\
    & - \frac{1}{{\rm Re}\,{\rm tr} \, \Sigma^{\prime LO}(-E_T)} \bigg[[\beta_1(\pv_+^2+\gamma_+^2)+\beta_2]\big(\big|{\mathcal A}_{(\ref{figX1a})}(\pv_+,m_+,m_0^*,-g/\sqrt{2}f_\pi)\big|^2  \\
    & - {\mathcal A}_{(\ref{figX1a})}(\pv_0,m_0,m_+^*,g/\sqrt{2}f_\pi){\mathcal A}_{(\ref{figX1a})}^*(\pv_+,m_+,m_0^*,-g/\sqrt{2}f_\pi)\big)  \\
    &+[\beta_3(\pv_0^2+\gamma_0^2)+\beta_4]\big(\big|{\mathcal A}_{(\ref{figX1a})}(\pv_0,m_0,m_+^*,g/\sqrt{2}f_\pi)\big|^2  \\
    & - {\mathcal A}_{(\ref{figX1a})}(\pv_+,m_+,m_0^*,-g/\sqrt{2}f_\pi){\mathcal A}_{(\ref{figX1a})}^*(\pv_0,m_0,m_+^*,g/\sqrt{2}f_\pi)\big) \bigg]  \\ 
    & - \frac{d\Gamma_0^{LO}(T_{cc}^+ \rightarrow D^+ D^0 \pi^0)}{d\pv_0^2d\pv_+^2} \frac{{\rm Re}\,\Sigma_0^{\prime NLO}}{{\rm Re}\,{\rm tr} \, \Sigma^{\prime LO}} \bigg|_{C_2 \rightarrow 0,E=-E_T}
\end{aligned}
\end{equation}

\begin{equation}
\begin{aligned}
    \frac{d\Gamma_0^{NLO}(T_{cc}^+\rightarrow D^0D^0\pi^+)}{d\pv_1^2d\pv_2^2} =&\; \frac{1}{{\rm Re}\,{\rm tr} \, \Sigma^{\prime LO}(-E_T)}{\rm Re}\, \bigg[ {\mathcal A}_{(\ref{figX1a})}(\pv_2,m_0,m_+^*,g/f_\pi)  \\
    & \times \bigg( \mathcal{A}_{(\ref{figX1b})}(\pv_1,m_0,m_0,m_{\pi^+},m_+^*,m_+^*,g/f_\pi,g/f_\pi,g/f_\pi)  \\
    & + {\mathcal A}_{(\ref{figX1b})}(\pv_2,m_0,m_0,m_{\pi^+},m_+^*,m_+^*,g/f_\pi,g/f_\pi,g/f_\pi)  \\
    & - {\mathcal A}_{(\ref{figX1b})}(\pv_1,m_+,m_0,m_{\pi^0},m_0^*,m_+^*,-g/\sqrt{2}f_\pi,g/\sqrt{2}f_\pi,g/f_\pi)  \\
    & - {\mathcal A}_{(\ref{figX1b})}(\pv_2,m_+,m_0,m_{\pi^0},m_0^*,m_+^*,-g/\sqrt{2}f_\pi,g/\sqrt{2}f_\pi,g/f_\pi)  \\
    & + {\mathcal A}_{(\ref{figX1d})}(\pv_1,m_0,m_0,m_{\pi^+},m_+^*,g/f_\pi,C_\pi^{(3)})  \\
    & - {\mathcal A}_{(\ref{figX1d})}(\pv_1,m_+,m_0,m_{\pi^0},m_0^*,-g/\sqrt{2}f_\pi,C_\pi^{(1)})  \\
    & +{\mathcal A}_{(\ref{figX1g})}(m_0,m_0,m_+^*,g/f_\pi,C_{0D}^{(1)}/2)\bigg)^* + (\pv_1 \leftrightarrow \pv_2)   \\
    & -\bigg(\frac{2g\mu_0}{f_\pi}\bigg)^2 \frac{\pv_\pi^2}{3} \beta_5\bigg(\frac{1}{\pv_1^2+\gamma_0^2}+\frac{1}{\pv_2^2+\gamma_0^2}\bigg)\bigg]  \\
    & - \frac{d\Gamma_0^{LO}(T_{cc}^+ \rightarrow D^0 D^0 \pi^+)}{d\pv_1^2d\pv_2^2}\bigg(\beta_4+\frac{{\rm Re}\,\Sigma_0^{\prime NLO}}{{\rm Re}\,{\rm tr} \,\Sigma^{\prime LO}} \bigg|_{C_2 \rightarrow 0,E=-E_T}\bigg)
\end{aligned}
\end{equation}
\end{widetext}

\begin{table*}
\caption{\label{tab:widths}Partial and total widths in units of keV at LO and NLO.}
\begin{ruledtabular}
\begin{tabular}{cccc}
& LO result & NLO lower bound & NLO upper bound \\
$\Gamma[T_{cc}^+\to D^0 D^0 \pi^+]$ & 28 & 21 & 44 \\
$\Gamma[T_{cc}^+\to D^+ D^0 \pi^0]$ & 13& 7.8 & 21 \\
$\Gamma_{\rm strong}[T_{cc}^+]$ & 41 & 29 & 66 \\
$\Gamma_{\rm strong}[T_{cc}^+]+\Gamma_{\rm EM}^{LO}[T_{cc}^+]$ & 47 & 35 & 72 \\ 
\end{tabular}
\end{ruledtabular}
\end{table*}

In the previous formulas we have used subscripts on $\mu$ and $\gamma$ to indicate which charm meson is a pseudoscalar in that particular channel, e.g., $\mu_0 = \mu(m_0,m_+^*)$. The combinations of self-energy diagrams that we need are ${\rm Re}\, {\rm tr}\,\Sigma^{\prime LO}(-E_T)$ and ${\rm Re}\, \Sigma_0^{\prime NLO}(-E_T,C_2\rightarrow 0)$.  In terms of the functions defined above, these are given by:
\begin{widetext}
\begin{equation}
\begin{aligned}
    {\rm Re}\, {\rm tr}\,\Sigma^{\prime LO} =&\; {\rm Re}\,\Sigma_1^\prime(m_0,m_+^*) + {\rm Re}\,\Sigma_1^\prime(m_+,m_0^*) \, ,  \\
    {\rm Re}\, \Sigma_0^{\prime NLO} |_{C_2\rightarrow 0} =&\; {\rm Re} \bigg[
    \Sigma_2^\prime(m_+,m_0^*,m_+,m_0^*,m_{\pi^+},g/f_\pi,g/f_\pi)  \\
    & + \Sigma_2^\prime(m_0,m_+^*,m_0,m_+^*,m_{\pi^+},g/f_\pi,g/f_\pi)  \\
    & + \Sigma_2^\prime(m_+,m_0^*,m_0,m_+^*,m_{\pi^0},-g/\sqrt{2}f_\pi,g/\sqrt{2}f_\pi)  \\
    & + \Sigma_2^\prime(m_0,m_+^*,m_+,m_0^*,m_{\pi^0},g/\sqrt{2}f_\pi,-g/\sqrt{2}f_\pi) \bigg]
\end{aligned}
\end{equation}
\end{widetext}

The expressions for $\beta_i$ are given in Appendix \ref{CpiAppendix}. 
The terms dependent on $\mathcal{A}_{{\rm(\ref{figX1b})}}$ and ${\rm Re}\, \Sigma_2^{\prime}$ have linear divergences that must cancel against each other.  They cancel exactly in the limit $\mu_0=\mu_+$.  We make that approximation in those terms only to ensure the cancellation; it is a reasonable approximation as $\mu_0/\mu_+ \approx 0.99948$.  See Appendix \ref{BasisInt} for more discussion of these linear divergences.

\section{Differential Decay Distributions and Partial Widths} \label{secPlots}

Once we have formulas for the $T_{cc}^+\rightarrow DD\pi$ partial widths, we can numerically integrate over part of three-body phase space in \emph{Mathematica} and plot the differential distribution $d\Gamma/dm_{DD}$.  It is insightful to compare our predicted curves to the LHCb experimental data for the total yield.  This will inform us about the effect and  importance of the different interactions in the effective theory.  We normalize our distributions by performing a least-squares fit of the LO distribution to the data, and using the same normalization factor for the NLO distributions. The $C_\pi$ decay diagrams, individually and as a whole, contribute negligibly to the distributions. The parameters $\beta_1$, $\beta_3$, and $\beta_5$ also have a small impact on the distributions over the range in which we vary them.  We therefore do not show plots varying these parameters individually.

\begin{figure*}[t]
\centering
\includegraphics[scale=0.8]{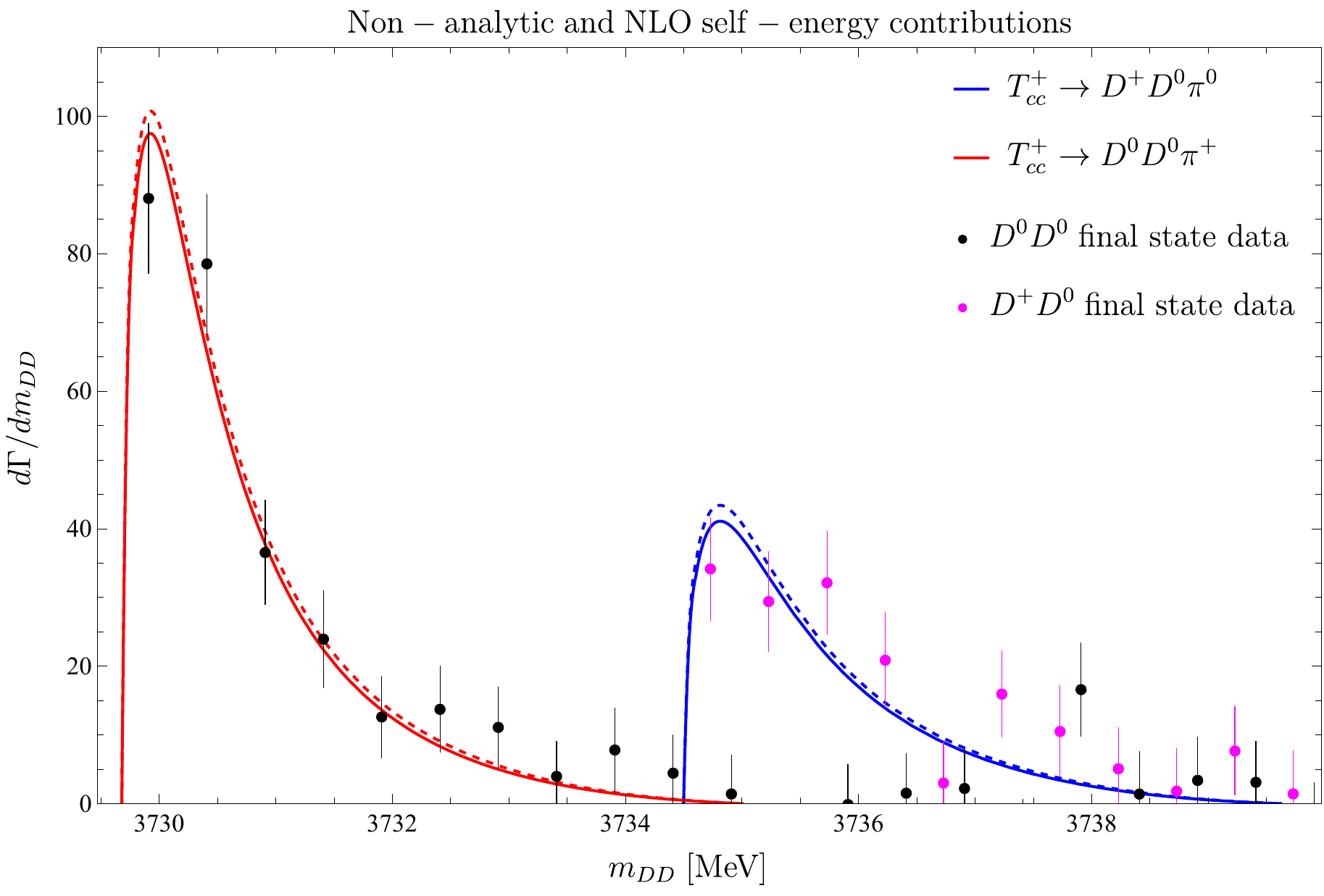}
\caption{A plot of the differential decay width as a function of the invariant mass of the final-state $D$ meson pair. Solid lines represent the LO calculation; the dashed lines represent the addition of non-analytic and NLO self-energy corrections.  Overlaid are the binned experimental data from LHCb, with the background subtracted.}
\label{ContributionsNAandSE}
\end{figure*}

The contributions from the non-$C_2$-dependent NLO self-energy corrections (i.e. the first diagram on the second line of Fig. \ref{bubbleDiagram}), as well as the contributions from Fig. \ref{figX1b}, serve to increase the partial widths by a small but noticeable amount (Fig. \ref{ContributionsNAandSE}).   The effect of the $C_{0D}$, $\beta_2$, and $\beta_4$ terms on the distributions can be significant.  In the following we will investigate their impact by setting all other contributions to $d\Gamma_{\rm NLO}/dm_{DD}$ to zero and varying them individually.  

\begin{figure*}[t]
\centering
\includegraphics[scale=0.8]{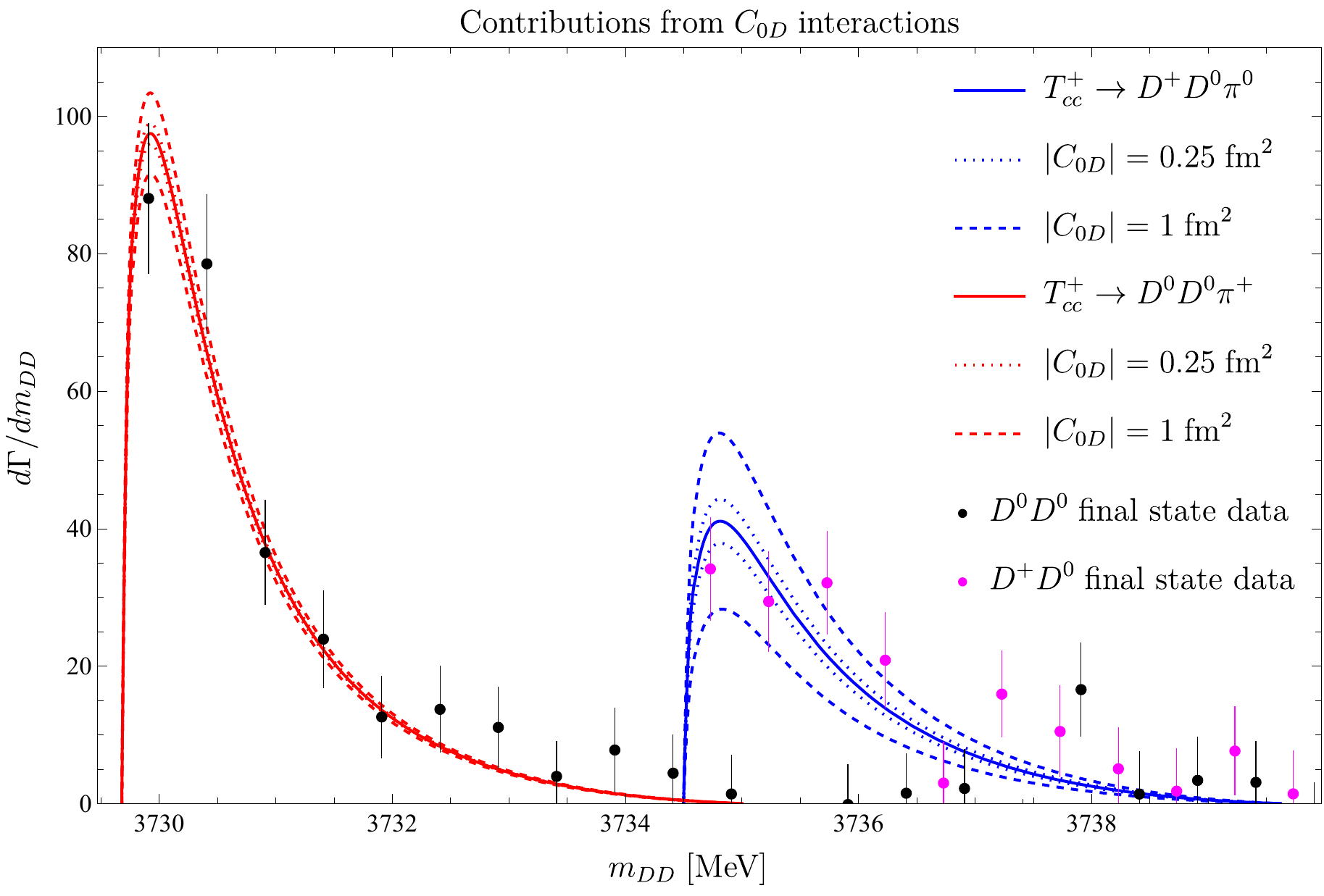}
\caption{A plot of the differential decay width as a function of the invariant mass of the final-state $D$ meson pair. Solid lines represent the LO calculation; The dashed and dotted lines represent two different ranges for $C_{0D}$.  Overlaid are the binned experimental data from LHCb, with the background subtracted.}
\label{ContributionsC0D}
\end{figure*}

The $C_{0D}$ interaction has a sizeable contribution to the partial widths, as evidenced in Fig. \ref{ContributionsC0D}, where we plot the differential distributions and vary this coupling in two possible ranges: $C_{0D} \in [-1,1]\; {\rm fm}^2$ and $\in [-0.25,0.25]\; {\rm fm}^2$.  Its effect on the neutral pion decay is twice as large as on the charged pion decay, because the coupling of charged pions to $D$ mesons is bigger by a factor of $\sqrt{2}$.   Clearly the differential distributions are sensitive to the coupling's magnitude.  If $C_{0D}$ is $+1\, {\rm fm}^2$ the peak of the$D^+D^0$ mass distribution is too high, and if it is $-1\, {\rm fm}^2$ three higher data points are underpredicted. It would be interesting to do a more careful analysis of the constraints these data put on $C_{0D}$ but that is beyond the scope of this paper. $C_{0D}$ is directly proportional to the $I=1$ $D$ meson scattering length, so more precise knowledge of $C_{0D}$ from lattice simulations or experiments would allow us to sharpen our predictions for  $T_{cc}^+$. 

\begin{figure*}[t]
\centering
\includegraphics[scale=0.8]{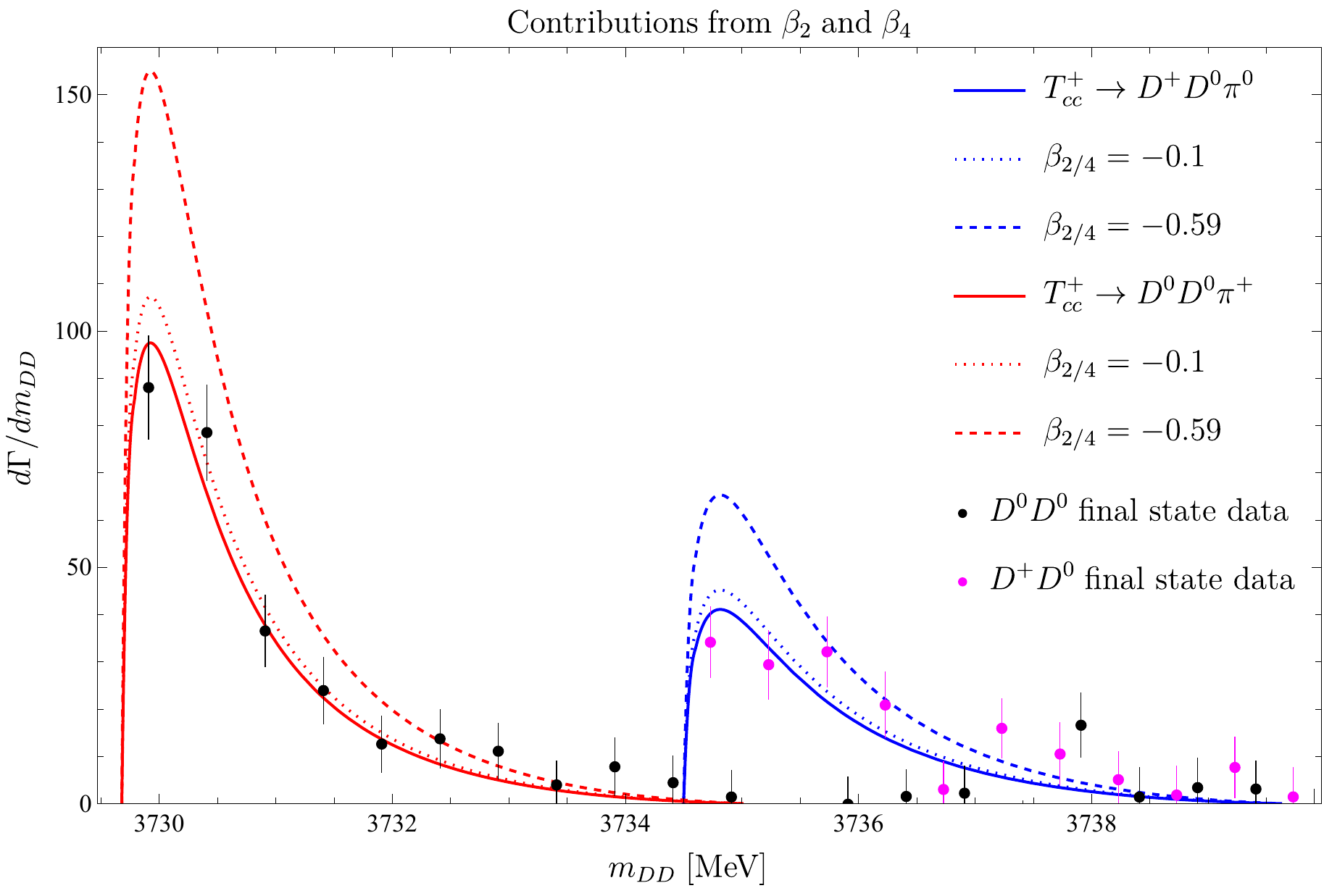}
\caption{A plot of the differential decay width as a function of the invariant mass of the final-state $D$ meson pair. Solid lines represent the LO calculation.  The dashed and dotted lines represent two different values of $\beta_{2}$ and $\beta_4$.  Overlaid are the binned experimental data from LHCb, with the background subtracted.}
\label{ContributionsBeta}
\end{figure*}

We can glean the significance of $\beta_2$ and $\beta_4$ by taking the isospin limit $m_0=m_+$.  In Appendix \ref{CpiAppendix} we see that in this limit:
\begin{equation}
\begin{aligned}
    \beta_2 = \beta_4 =-\gamma r_0 \, ,
\end{aligned}
\end{equation}
where $\gamma$ is the binding momentum and $r_0$ is the effective range in the $I=0$ channel. The effective range is positive and we expect $\gamma r_0 < 1$. In Fig. \ref{ContributionsBeta}, we plot the distribution with all other NLO interactions turned off, and for two values of $\beta_2=\beta_4\equiv \beta$: $-0.1$ and $-0.59$, along with the LO curve ($\beta=0$). 
We get $\gamma r_0 = 0.59$ if we use the largest 
binding momentum $(\gamma_+)$ and $r_0=1/(100\, {\rm MeV})$. For nucleons, $r_0 \approx 1/(100\, {\rm MeV})$; since charm mesons are considerably more compact objects one might expect the effective range for charm mesons to be smaller.
We can see that the distribution is highly sensitive to  the choice of $\beta$.  A $\beta$ of $-0.59$  greatly increases the differential distribution, and is in much poorer agreement with the experimental data.  This suggests that the effective range for $T_{cc}^+$ is smaller than for nucleons.  

Clearly the partial widths and their differential distributions can vary substantially depending on the choice of parameters in the effective field theory.  However, the availability of experimental data for the decays presents the possibility of performing fits of the distributions to the data to obtain estimates for these parameters.  This could improve the predictive power of the effective theory.  We save such a careful statistical analysis for a future publication.

\begin{figure*}[t]
\centering
\includegraphics[scale=0.8]{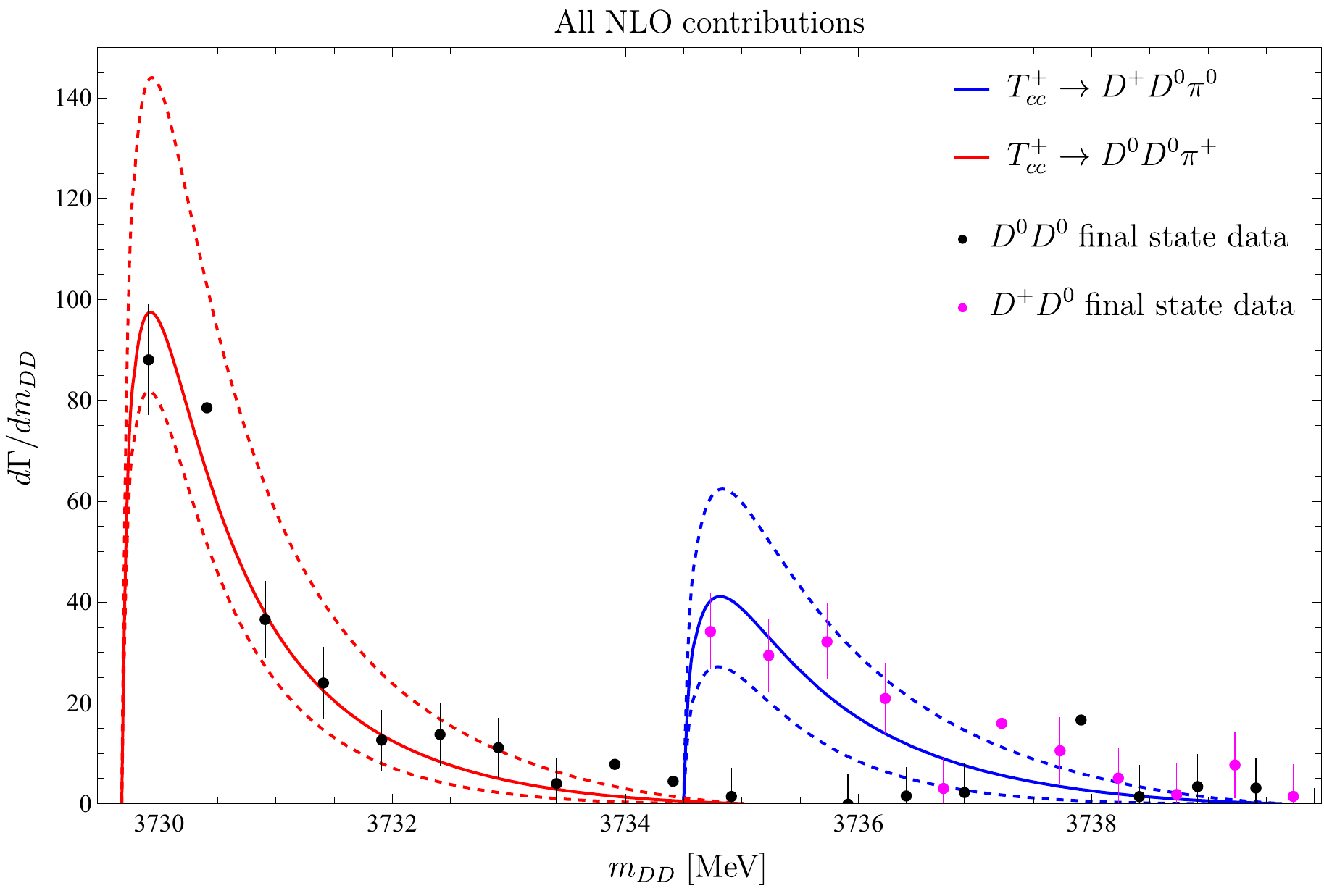}
\caption{A plot of the differential decay width as a function of the invariant mass of the final-state $D$ meson pair. Solid lines represent LO calculation; the dashed lines represent the lower and upper bounds of the NLO corrections.  Here, we vary $-1 \; {\rm fm}^2 \leq C_{0D} \leq 0.25 \; {\rm fm}^2$ and $-0.26\leq \beta_{2/4} \leq 0$. Overlaid are the binned experimental data from LHCb, with the background subtracted.}
\label{ContributionsAll}
\end{figure*}

We can use these plots that show the effect of a subset of the NLO contributions to inform which ranges for the parameters to use when estimating the total NLO contribution to the differential distribution (Fig. \ref{ContributionsAll}). The upper and lower bounds in the figure reflect varying $C_{0D}$ from $-1\; {\rm fm}^2$ to $0.25\; {\rm fm}^2$. The parameters $\beta_1$, $\beta_3$, and $\beta_5$ are varied from $-1/(100\, {\rm MeV})^2$ to $+1/(100\, {\rm MeV})^2$.  The parameters $\beta_2$ and $\beta_4$, which reduce to $-\gamma r_0$ in the isospin limit, are varied between $0$ and $-0.26$.  The latter value corresponds to a binding momentum for the $D^{*+}D^0$ channel, $\gamma_0$, and $r_0=1/(100\, {\rm MeV})$. While the uncertainty in the total width of the $T_{cc}^+$ can be significant depending on the values of the NLO couplings, the qualitative aspects of the plots of the differential decay widths in Fig. \ref{ContributionsAll} are consistent between LO and NLO. The overall shape and location of the peaks are unchanged by pion exchange and final-state rescattering. 

When integrating over the full phase space to get the partial widths, we use the same ranges for the parameters as in Fig. \ref{ContributionsAll}.  The partial widths are given in Table \ref{table:widths}. Note that the LO numbers differ from those in our original paper \cite{Fleming:2021wmk} because here we use the binding energy from the unitarized Breit-Wigner fit, whereas in Ref.~\cite{Fleming:2021wmk} we used the value from the $P$-wave two-body Breit Wigner fit with a Blatt-Weisskopf form factor. This has the effect of slightly increasing the prediction for the width compared to the initial paper, bringing it closer to the experimental value. When adding the LO electromagnetic decay width of $6.1 \, {\rm keV}$ (which is only slightly affected by the different binding energy) the total LO width predicted by our effective theory is $47 \, {\rm keV}$ which is already in excellent agreement with the LHCb experimental value of $48 \, {\rm keV}$.  Adding in the NLO contribution to the strong decay widths, the total width of the $T_{cc}^+$ can range from $35 \, {\rm keV}$ to $72 \, {\rm keV}$. So we can establish an uncertainty in the width due to NLO strong decays of $\Gamma[T_{cc}^+] = 47_{-25\%}^{+53\%}\,{\rm keV}$. This is comparable to the uncertainty from similar operators contributing to the decay of $\chi_{c1}(3872)$ in XEFT \cite{Dai:2019hrf}.

We did not consider NLO corrections to the electromagnetic decay, because the LO electromagnetic decay was already a small contribution to the total width.  In particular, the differential distribution for the electromagnetic decay was negligible compared to the strong decays' distributions.

\begin{figure*}[t]
\centering
\includegraphics[scale=0.8]{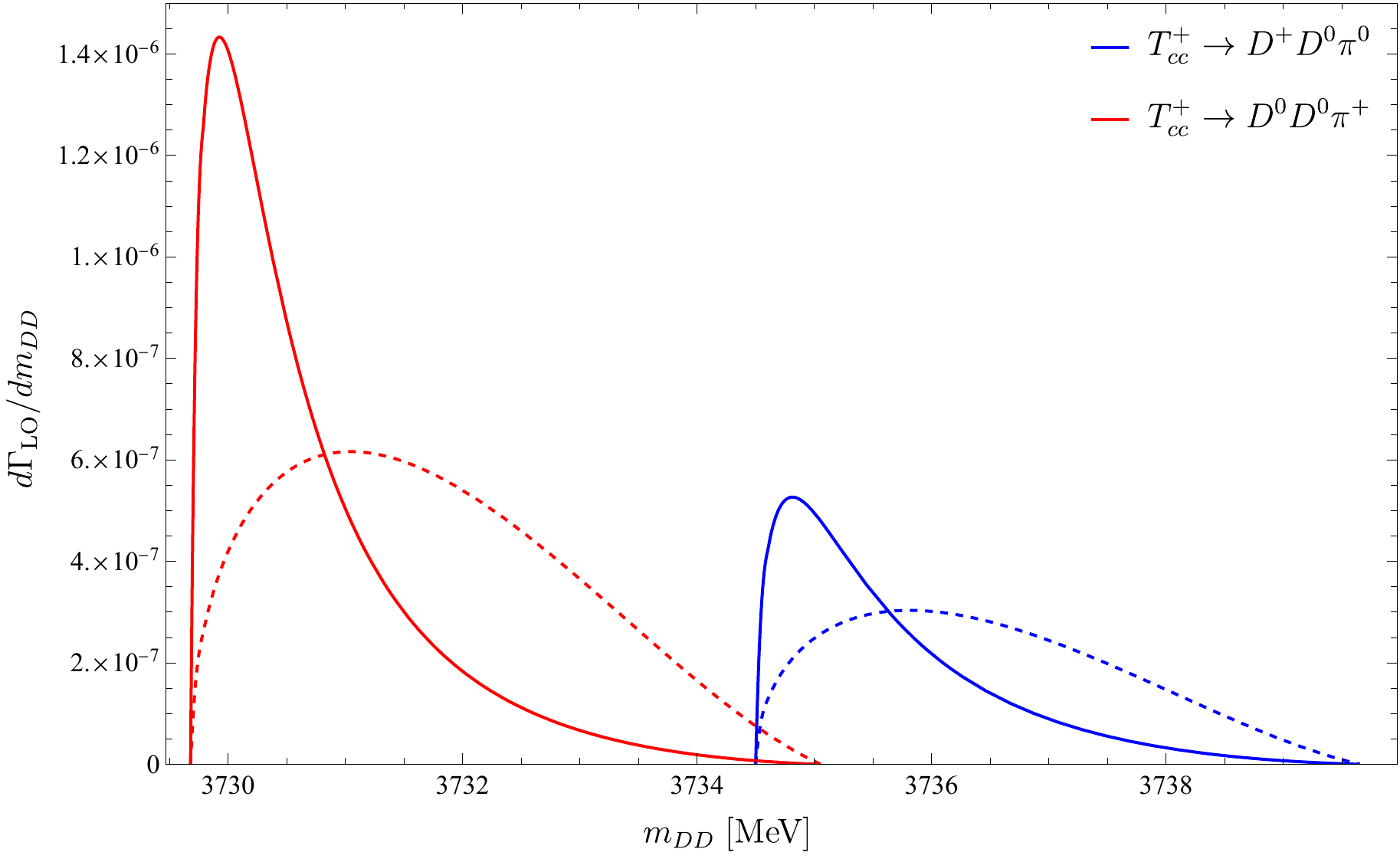}
\caption{Comparing our LO differential decay width to one where the $D^*$ propagators are taken to be constant.  The curves are fixed to have the same normalization. Note the lack of a sharp peak in the constant propagator curves.}
\label{constantProp}
\end{figure*}

To illustrate why these differential decay width plots are good tests of the molecular nature of the $T_{cc}^+$, in Fig. \ref{constantProp} we can compare the LO differential curves to those which would arise if we replaced the virtual $D^*$ propagators with a constant. The latter do not have sharp peaks and thus would be in poor agreement with the experimental data.

\section{Conclusions} \label{secConclusions}

In this paper we have determined the effects of NLO strong decays on the total width and differential decay width of the exotic meson $T_{cc}^+$.  We considered pion exchange and final-state rescattering diagrams, from similar operators to those in XEFT for the $\chi_{c1}(3872)$ \cite{Dai:2019hrf}. We arrive at similar conclusions as Ref.~\cite{Dai:2019hrf}. The differential decay width plots have shapes and  peaks that are relatively unchanged by the NLO effects, but the total width has significant  uncertainty: $\Gamma[T_{cc}^+] = 47_{-25\%}^{+53\%}\,{\rm keV}$.  The central value (the LO result) is in good agreement with data.

We varied the parameters in the NLO calculation to get a sense of the uncertainty in the predictions and determine which parameters in the NLO calculation give the biggest corrections. Nonanalytic corrections for pion loops are not important. The  parameter $C_{0D}$, which is proportional to the $I=1$ $D$ meson scattering length, and  $\beta_{2}$ and $\beta_4$, which in the isospin limit are equal and proportional to the $I=0$ $D$ meson effective ranges, significantly affect the decay width and normalization of the differential distribution.    It would be interesting to fit the NLO differential curves to the experimental data and obtain bounds on the undetermined couplings, thereby learning more about these physical quantities. Alternatively, one might hope to get information about these parameters from lattice simulations or other experiments.  Any improvement in our understanding of these parameters in $D$ meson scattering would increase the predictive power of the effective field theory.

{\bf Acknowledgments} - L.~D. is supported by the Alexander von Humboldt Foundation. S.~F. is supported by the U.S. Department of Energy, Office of Science, Office of Nuclear Physics,  under award number DE-FG02-04ER41338. T.~M. and R.~H. are supported by the U.S. Department of Energy, Office of Science, Office of Nuclear Physics under grant Contract Numbers  DE-FG02-05ER41367.

\appendix

\section{coupled-channel decay width} \label{CoupledChannelAppendix}

The full expression for the isospin 0 two-point correlator is
\begin{equation}
\begin{aligned}
-iG_0 =&\; \frac{-\frac{1}{2}\Sigma_0-2C_0^{(1)}{\rm det}\,\Sigma}{1+C_0^{(0)}\Sigma_0+C_0^{(1)}\Sigma_1+4C_0^{(0)}C_0^{(1)}{\rm det}\,\Sigma} \, ,
\end{aligned}
\end{equation}
where $\Sigma_{0/1} \equiv \Sigma_{11}+\Sigma_{22}\mp\Sigma_{12}\mp\Sigma_{21}$ are the isospin 0 and isopsin-1 combinations of the elements of $\Sigma$. Since we expect $T_{cc}^+$ to be an isospin 0 state we treat $C_0^{(1)}$ perturbatively and expand to NLO in $C_0^{(1)}$.
\begin{equation}
\begin{aligned}
-iG_0 \approx & \; \frac{1}{2}\frac{-\Sigma_0}{1+C_0^{(0)}\Sigma_0} + \frac{1}{2}\frac{C_0^{(1)}(\Sigma_{11}^{LO}-\Sigma_{22}^{LO})^2}{(1+C_0^{(0)}\Sigma_0)^2} \, .
\end{aligned}
\end{equation}
We see that the real numerator of the $C_0^{(1)}$ term is the residue of a double pole at $1+C_0^{(0)}\Sigma_0=0$.  That can be interpreted physically as a small shift in the location of the bound state, which can be seen from expanding the right-hand side of Eq.~(\ref{G01}) about $E_T^{NLO} = E_T - E_T^{LO}$.  But since we are already tuning $E_T$ to be the location of the $T_{cc}^+$ bound state, we can set $C_0^{(1)}$ to zero to remove the double pole from the amplitude.
\begin{equation}
\begin{aligned}
-iG_0 \rightarrow & \; \frac{1}{2}\frac{-\Sigma_0}{1+C_0^{(0)}\Sigma_0} \, .
\end{aligned}
\end{equation}
At this stage the problem is identical to the single-channel problem in XEFT \cite{Fleming:2007rp}, with the single-channel two-point function replaced by our isospin 0 combination of coupled-channel two-point functions.  The wave function renormalization and decay width are therefore:
\begin{equation}
\begin{aligned}
Z_0 =&\; \frac{1}{\big(C_0^{(0)}\big)^2 {\rm Re}\, \Sigma_0^\prime(-E_T)} \, ,  \\
\Gamma_0 =&\; \frac{2\, {\rm Im}\, \Sigma_0(-E_T)}{{\rm Re}\, \Sigma_0^\prime(-E_T)} \, .
\end{aligned}
\end{equation}
$\Sigma_0$ has LO contributions from the diagonal elements, and NLO contributions from all elements.  After expanding in the NLO terms we find our corrections to the LO decay width.  
\begin{equation} \label{width2}
\begin{aligned}
\Gamma_0  \approx & \; \Gamma^{LO}\bigg(1-\frac{{\rm Re}\, \Sigma_0^{\prime NLO}(-E_T)}{{\rm Re}\, {\rm tr}\,\Sigma^{\prime LO}(-E_T)}\bigg)  \\
& + \frac{2\,{\rm Im}\, \Sigma_0^{NLO}(-E_T)}{{\rm Re}\, {\rm tr}\,\Sigma^{\prime LO}(-E_T)} \, .
\end{aligned}
\end{equation}

\section{Basis integrals and the PDS scheme} \label{BasisInt}

The most basic integral that arises when evaluating the one-loop diagrams in the PDS scheme is:
\begin{widetext}
\begin{equation} \label{PDSint}
\begin{aligned}
\bigg(\frac{\Lambda_{\rm PDS}}{2}\bigg)^{4-d} \int \frac{d^{d-1}{\bf l}}{(2\pi)^{d-1}} \frac{1}{\lv^2+c-i\epsilon} = & \; \frac{1}{4\pi}(\Lambda_{\rm PDS}-\sqrt{c-i\epsilon}) \, .
\end{aligned}
\end{equation}
\end{widetext}
This result is obtained by subtracting the pole in $d=3$ with a counterterm, then evaluating the result in $d=4$, yielding a linear divergence in $\Lambda_{\rm PDS}$. 

The scalar integral $I(\pv)$ is finite in $d=3$ and $d=4$, so no PDS counterterm is needed.
\begin{widetext}
\begin{equation}
\begin{aligned}
I({\bf p}) =&\; \int \frac{d^{d-1}{\bf l}}{(2\pi)^{d-1}} \frac{1}{{\bf l}^2+c_1-i\epsilon} \frac{1}{{\bf l}^2-2b{\bf l}\cdot{\bf p}+c_2-i\epsilon}  \\
=&\; \frac{1}{8\pi}\frac{1}{\sqrt{b^2\pv^2}}\bigg[\tan^{-1}\bigg( \frac{c_2-c_1}{2\sqrt{b^2\pv^2c_1}}\bigg) + \tan^{-1}\bigg(\frac{2b^2\pv^2+c_1-c_2}{2\sqrt{b^2\pv^2(c_2-b^2\pv^2)}} \bigg) \bigg] \, .
\end{aligned}
\end{equation}
\end{widetext}

The linear tensor integral $I^{(1)}(\pv)$ can be solved using algebraic manipulation of the numerator, which yields two integrals of the form of Eq.~(\ref{PDSint}) that have opposite sign for the divergence, and so $I^{(1)}(\pv)$ is UV finite.
\begin{widetext}
\begin{equation}
\begin{aligned}
\pv^iI^{(1)}({\bf p}) =&\; \int \frac{d^{d-1}{\bf l}}{(2\pi)^{d-1}} \lv^i \frac{1}{{\bf l}^2+c_1-i\epsilon} \frac{1}{{\bf l}^2-2b{\bf l}\cdot{\bf p}+c_2-i\epsilon} \, ,  \\
\rightarrow \pv^2I^{(1)}({\bf p})=&\; \frac{1}{2b}\bigg[\frac{1}{4\pi}\sqrt{c_1-i\epsilon}-\frac{1}{4\pi}\sqrt{c_2-b^2\pv^2-i\epsilon}+(c_2-c_1)I(\pv)\bigg] \, .
\end{aligned}
\end{equation}
\end{widetext}

The quadratic tensor integrals $I^{(2)}$ require care when implementing the PDS scheme.  The linear divergences which arise in the decay width can only cancel if the subtraction scheme is implemented correctly.  After using Feynman parameters to combine the propagators and obtain an integrand like $\lv^i\lv^j f(\lv^2)$, the correct procedure is to replace $\lv^i\lv^j\rightarrow \delta^{ij}/3$ immediately, and not with $\delta^{ij}/(d-1)$.  The latter would cancel the factor of $d-1$ that arises when evaluating the loop momentum integral, and this results in the incorrect coefficient for the PDS subtraction scale $\Lambda_{\rm PDS}$.  Additionally, algebraic manipulation of the numerator of $I^{(2)}$ to reduce it to integrals of the form of $I^{(1)}$ and $I$ leads to yet another incorrect coefficient.  This is the method used to obtain the expressions in the appendix of Ref.~\cite{Dai:2019hrf}; as such, the formulas for the decay width in that paper are only correct if $\Lambda_{\rm PDS}=0$ and $d=4$.  

Using the correct procedure for the basis integrals gives the following results:
\begin{widetext}
\begin{equation}
\begin{aligned}
\pv^i\pv^jI^{(2)}_0({\bf p})+\delta^{ij}\pv^2 I^{(2)}_1({\bf p}) =&\; \int \frac{d^{d-1}{\bf l}}{(2\pi)^{d-1}} \lv^i \lv^j \frac{1}{{\bf l}^2+c_1-i\epsilon} \frac{1}{{\bf l}^2-2b{\bf l}\cdot{\bf p}+c_2-i\epsilon} \, ,  \\
I_0^{(2)}(\pv) =&\; \frac{b^2}{8\pi} \int_0^1 dx \, \frac{x^2}{\sqrt{\Delta (x)}} \, , \\
\rightarrow\pv^2I_1^{(2)}(\pv) =&\; \frac{1}{8\pi}\bigg[\frac{2}{3}\Lambda_{\rm PDS} - \int_0^1 dx\, \sqrt{\Delta (x)}\bigg] \, , 
\end{aligned}
\end{equation}
\end{widetext}
for $\Delta (x)=-b^2\pv^2 x^2+(c_2-c_1)x+c_1-i\epsilon$.  One can be reassured that this implementation of the PDS scheme is correct because the same relative weight of the $\Lambda_{\rm PDS}$ and $\int_0^1 dx\, \sqrt{\Delta (x)}$ terms is obtained when using a hard cutoff.  That does not occur when using $\lv^i \lv^j \rightarrow \delta^{ij}/(d-1)$ or algebraic manipulation of the numerator.  Furthermore, unless the relative weight of the cutoff-dependent terms in $I_1^{(2)}$ and ${\rm Re} \, \Sigma_2^\prime$ is $2/3$, the linear divergences that appear in $\Gamma_0^{\rm NLO}$ as $\mathcal{A}_{(\ref{figX1b})}$ and ${\rm Re} \, \Sigma_2^\prime$ do not cancel in the isospin limit, as they do in XEFT.  For the $T_{cc}^+$, they cancel when $\mu_0=\mu_+$, an approximation we make in the cutoff-dependent terms to ensure cancellation.

With algebraic manipulation of the integrand in $I_0^{(2)}$ and integration by parts in $I_1^{(2)}$, we can rewrite these expressions in terms of $I$ and $I^{(1)}$.
\begin{equation}
\begin{aligned}
\pv^2 I_0^{(2)} =&\; -\frac{1}{16\pi}\sqrt{c_2-b^2\pv^2-i\epsilon} + \frac{c_1}{2}I(\pv)  \\ 
&+\frac{3}{4}\frac{c_2-c_1}{b}I^{(1)}(\pv) \, , \\
\pv^2 I_1^{(2)} =&\; \frac{\Lambda_{\rm PDS}}{12\pi} -\frac{1}{16\pi}\sqrt{c_2-b^2\pv^2-i\epsilon}  \\ 
& - \frac{c_1}{2}I(\pv)-\frac{1}{4}\frac{c_2-c_1}{b}I^{(1)}(\pv) \, .
\end{aligned}
\end{equation}

\section{$C_\pi$ couplings and $\beta_i$ expressions} \label{CpiAppendix}

In the isospin $\ket{I,m_I}$ basis, we use the phase convention
\begin{equation}
\begin{aligned}
    \ket{\pi^+} =&\; -\ket{1,1} \, , \quad \ket{\pi^0} = \ket{1,0} \, ,  \\
    \ket{D^+} =&\; \Ket{\frac{1}{2},\frac{1}{2}} \, , \quad \ket{D^0} = \Ket{\frac{1}{2},-\frac{1}{2}} \, . 
\end{aligned}
\end{equation}
Then the Clebsch-Gordan decomposition of the $D\pi$ pairs is
\begin{align}
    \ket{D^0\pi^0} =&\; \sqrt{\frac{2}{3}}\Ket{\frac{3}{2},-\frac{1}{2}} + \frac{1}{\sqrt{3}}\Ket{\frac{1}{2},-\frac{1}{2}} \, ,  \nonumber \\
    \ket{D^+\pi^0} =&\; \sqrt{\frac{2}{3}}\Ket{\frac{3}{2},\frac{1}{2}} + \frac{1}{\sqrt{3}}\Ket{\frac{1}{2},\frac{1}{2}} \, ,  \\
    \ket{D^0\pi^+} =&\; -\sqrt{\frac{2}{3}}\Ket{\frac{1}{2},\frac{1}{2}} - \frac{1}{\sqrt{3}}\Ket{\frac{3}{2},\frac{1}{2}} \nonumber \, .
\end{align}
From this we can deduce
\begin{equation} \label{scattlen}
\begin{aligned}
    a_{D^0\pi^0} =&\; a_{D^+\pi^0} = \frac{2}{3}a_{D_\pi}^{3/2}+\frac{1}{3}a_{D\pi}^{1/2} \, ,  \\
    a_{D^0\pi^+} =&\; \frac{1}{3}a_{D\pi}^{3/2}+\frac{2}{3}a_{D\pi}^{1/2} \, .
\end{aligned}
\end{equation}
These scattering lengths are calculated on the lattice in Ref.~\cite{Liu:2012zya} to be $a_{D\pi}^{1/2} = 0.37_{-0.02}^{+0.03}\, {\rm fm}$ and $a_{D\pi}^{3/2} = -(0.100\pm0.002)\, {\rm fm}$.  The matching from tree-level scattering tells us that, for the diagonal couplings $C_\pi^{(2)}$ and $C_\pi^{(3)}$, we can use $C_\pi = 4\pi(1+m_\pi/m_D)a_{D\pi}$, with the appropriate masses and scattering lengths for each process. We can then use those two values to solve for $C_\pi^{(1/2)}$ and $C_\pi^{(3/2)}$ and obtain $C_\pi^{(1)}$.  We get
\begin{equation}
\begin{aligned}
    C_\pi^{(1)} =&\; -3.0_{-0.40}^{+0.32}\, {\rm fm} \, ,  \\
    C_\pi^{(2)} =&\; -0.76_{-0.09}^{+0.14}\, {\rm fm} \, ,  \\
    C_\pi^{(3)} =&\; 2.9_{-0.2}^{+0.3}\, {\rm fm} \, . 
\end{aligned}
\end{equation}
The expressions for the $\beta_i$ are given below.  The subscripts on the $\gamma$ and $\mu$ variables indicate the pseudoscalar charm meson is in that channel, e.g. $\gamma_+ = \gamma(m_+,m_0^*)$ is the binding momentum in the channel with the $D^+$ meson.
\begin{widetext}
\begin{equation}
\begin{aligned}
    \beta_1 =&\; (\Lambda_{\rm PDS}-\gamma_+ )\bigg(\frac{f_\pi}{\sqrt{2}\pi g}B_1^{(1)}+\frac{1}{\pi}C_2^{(+)}\mu_+-\frac{1}{\pi}C_2^{(-)}\mu_0 \frac{\Lambda_{\rm PDS}-\gamma_0}{\Lambda_{\rm PDS}-\gamma_+} \bigg) \, , \\
\end{aligned}
\end{equation}
\begin{equation}
\begin{aligned}
    \beta_2 =&\; \bigg[\frac{1}{\pi}C_2^{(+)}\mu_+(-2\gamma_+^2)(\Lambda_{\rm PDS}-\gamma_+)-\frac{1}{\pi}C_2^{(-)}\mu_0(-\gamma_0^2-\gamma_+^2)(\Lambda_{\rm PDS}-\gamma_0)  \\
    & +2\pi\bigg(\frac{\mu_0^2}{\gamma_0}+\frac{\mu_+^2}{\gamma_+}\bigg)^{-1}\bigg[-\frac{1}{\pi^2}C_2^{(+)}\mu_+^3(\gamma_+-\Lambda_{\rm PDS})(2\gamma_+-\Lambda_{\rm PDS})  \\
    & -\frac{1}{\pi^2}C_2^{(+)}\mu_0^3(\gamma_0-\Lambda_{\rm PDS})(2\gamma_0-\Lambda_{\rm PDS})  \\
    & -\frac{C_2^{(-)}(\gamma_+^2+\gamma_0^2)\mu_+\mu_0}{2\pi}\bigg(\frac{\mu_+}{\gamma_0}(\Lambda_{\rm PDS}-\gamma_0)+\frac{\mu_0}{\gamma_+}(\Lambda_{\rm PDS}-\gamma_+)\bigg)  \\
    & +\frac{C_2^{(-)}\mu_+\mu_0(\mu_++\mu_0)}{\pi^2}(\Lambda_{\rm PDS}-\gamma_+)(\Lambda_{\rm PDS}-\gamma_0)\bigg]\bigg] \, , \\
\end{aligned}
\end{equation}
\begin{equation}
\begin{aligned}
    \beta_3 =&\; (\Lambda_{\rm PDS}-\gamma_0)\bigg(-\frac{f_\pi}{\sqrt{2}\pi g}B_1^{(2)}+\frac{1}{\pi}C_2^{(+)}\mu_0-\frac{1}{\pi}C_2^{(-)}\mu_+ \frac{\Lambda_{\rm PDS}-\gamma_+}{\Lambda_{\rm PDS}-\gamma_0} \bigg) \, , \\
\end{aligned}
\end{equation}
\begin{equation}
\begin{aligned}
    \beta_4 =&\; \bigg[\frac{1}{\pi}C_2^{(+)}\mu_0(-2\gamma_0^2)(\Lambda_{\rm PDS}-\gamma_0)-\frac{1}{\pi}C_2^{(-)}\mu_+(-\gamma_0^2-\gamma_+^2)(\Lambda_{\rm PDS}-\gamma_+)  \\
    & +2\pi\bigg(\frac{\mu_0^2}{\gamma_0}+\frac{\mu_+^2}{\gamma_+}\bigg)^{-1}\bigg[-\frac{1}{\pi^2}C_2^{(+)}\mu_+^3(\gamma_+-\Lambda_{\rm PDS})(2\gamma_+-\Lambda_{\rm PDS})  \\
    & -\frac{1}{\pi^2}C_2^{(+)}\mu_0^3(\gamma_0-\Lambda_{\rm PDS})(2\gamma_0-\Lambda_{\rm PDS})  \\
    &-\frac{C_2^{(-)}(\gamma_+^2+\gamma_0^2)\mu_+\mu_0}{2\pi}\bigg(\frac{\mu_+}{\gamma_0}(\Lambda_{\rm PDS}-\gamma_0)+\frac{\mu_0}{\gamma_+}(\Lambda_{\rm PDS}-\gamma_+)\bigg)  \\
    & +\frac{C_2^{(-)}\mu_+\mu_0(\mu_++\mu_0)}{\pi^2}(\Lambda_{\rm PDS}-\gamma_+)(\Lambda_{\rm PDS}-\gamma_0)\bigg]\bigg] \, , \\
\end{aligned}
\end{equation}
\begin{equation}
\begin{aligned}
    \beta_5 =&\; \frac{1}{\pi}C_2^{(+)}\mu_0(\Lambda_{\rm PDS}-\gamma_0)-\frac{1}{\pi}C_2^{(-)}\mu_+(\Lambda_{\rm PDS}-\gamma_+)  \\
    & +\frac{B_1^{(3)}f_\pi}{4\pi g}(\gamma_0-\Lambda_{\rm PDS})-\frac{B_1^{(4)}f_\pi}{4\pi g}(\gamma_+-\Lambda_{\rm PDS})\frac{\mu_+}{\mu_0} \, . 
\end{aligned}
\end{equation}
\end{widetext}
It is instructive to take the isospin limit of these $\beta$ expressions and compare to XEFT.  Referring to Eq.~(\ref{LagB11}), we can write down the $B_1$ couplings in this limit.
\begin{equation}
\begin{aligned}
    B_1^{(1)} =&\; - B_1^{(2)} = -\sqrt{2}B_1^{(I=0)} \, ,  \\
    B_1^{(3)} =&\; 2(B_1^{(I=1)}+B_1^{(I=0)}) \, ,  \\
    B_1^{(4)} =&\; 2(B_1^{(I=1)}-B_1^{(I=0)}) \, . 
\end{aligned}
\end{equation}
Then taking $\mu_+=\mu_0=\mu$, $\gamma_+=\gamma_0 = \gamma$ we find:
\begin{equation}
\begin{aligned}
    \beta_1 = \beta_3=\beta_5 = & \; \frac{1}{\pi}(\gamma-\Lambda_{\rm PDS}) \\
    & \times \bigg(\frac{B_1^{(I=0)}f_\pi}{g}-2C_2^{(0)}\mu\bigg) \, ,  \\
    \beta_2 = \beta_4 = & \; -\frac{4C_2^{(0)}\mu\gamma}{\pi}(\gamma-\Lambda_{\rm PDS})^2 \, .
\end{aligned}
\end{equation}
The isospin 1 couplings drop out, which is to be expected given that we have projected out the isospin 0 state and are here dropping isospin-breaking interactions.  These expressions also match the dependence of the decay rate on $C_2$ and $B_1$ in XEFT \cite{Fleming:2007rp}.  Using Eq.~(24) of \cite{Fleming:2007rp} (and adjusting for a factor of 4 in the definition of $C_2$ in that paper) we see that $\beta_2=\beta_4=-\gamma r_0$ in the isospin limit. It is an important check on our calculation that  in the isospin limit the theory can be properly renormalized with isospin-respecting counterterms. When isospin breaking in the masses and binding momentum is included, isospin breaking in the $B_1$ operators needs to be included as we have done in this paper.


\bibliography{main}

\end{document}